%% file: prl5.tex
\RequirePackage{lineno}
\documentclass[aps,prl,twocolumn,showpacs,superscriptaddress,showpacs]{revtex4}
\usepackage{epsfig}
\usepackage{graphicx} 
\usepackage{dcolumn}  
\usepackage{bm}       
\usepackage{amssymb}  

\def\mevc {\ifmmode {\rm MeV}/c \else MeV$/c$\fi}
\def\mevcc {\ifmmode {\rm MeV}/c^2 \else MeV$/c^2$\fi}
\def\gevc {\ifmmode {\rm GeV}/c \else GeV$/c$\fi}
\def\gevcc {\ifmmode {\rm GeV}/c^2 \else GeV$/c^2$\fi}
\def\tevcc {\ifmmode {\rm TeV}/c^2 \else TeV$/c^2$\fi}

\def\ol   {\overline}

\def\vtd  {\ifmmode |V_{td}| \else $|V_{td}|$\fi}
\def\vtb  {\ifmmode |V_{tb}| \else $|V_{tb}|$\fi}
\def\vts  {\ifmmode |V_{ts}| \else $|V_{ts}|$\fi}
\def\vcb  {\ifmmode |V_{cb}| \else $|V_{cb}|$\fi}

\newcommand{\Ds} {\ifmmode D_{\mbox{\sl s}}^{-}
                       \else $D_{\mbox{\sl s}}^{-}$\fi}
\newcommand{\Bs} {\ifmmode B_{\mbox{\sl s}}^{0}
                       \else $B_{\mbox{\sl s}}^{0}$\fi}
\newcommand{\Bsb} {\ifmmode \ol B_{\mbox{\sl s}}^{0}
                       \else $\ol B_{\mbox{\sl s}}^{0}$\fi}
\newcommand{\Bsh} {\ifmmode B_{\mbox{\sl s}}^H
                       \else $B_{\mbox{\sl s}}^H$\fi}
\newcommand{\Bsl} {\ifmmode B_{\mbox{\sl s}}^L
                       \else $B_{\mbox{\sl s}}^L$\fi}
\newcommand{\Dsl} {\ifmmode D_{\mbox{\sl s}}^{-} \ell^+
                       \else $D_{\mbox{\sl s}}^{-} \ell^+$\fi}
\newcommand{\xs} {\ifmmode x_{\mbox{\sl s}}
                       \else $x_{\mbox{\sl s}}$\fi}
\newcommand{\xd} {\ifmmode x_d \else $x_d$\fi}
\newcommand{\lxy} {\ifmmode L_{\rm xy} \else $L_{\rm xy}$\fi}
\newcommand{\dgam} {\ifmmode \Delta\Gamma \else $\Delta\Gamma$\fi}
\newcommand{\dm} {\ifmmode \Delta m \else $\Delta m$\fi}
\newcommand{\ctau} {\ifmmode c\tau \else $c\tau$\fi}
 
\newcommand{\et}{{ E}_{\mathrm{T}}}

\newcommand{\abseta}{|\eta| \leq}

\newcommand{\met}{\mbox{$\protect \raisebox{.3ex}{$\not$}\et$}}

\newcommand{\ppbar}{ { p} \bar{{ p}} }

\newcommand{\ttbar}{t\bar{t}}

\newcommand{\tbbar} { { t} \bar{{b}}  }

\newcommand{\ifb}{ {\rm fb}^{-1} }

\begin{document}
\bibliographystyle{apsrev}


\title{Search for the Production of Narrow \boldmath${\tbbar}$ Resonances in 1.9\,$\ifb$ of \boldmath${\ppbar}$ Collisions at \boldmath${\sqrt{s}=}$ 1.96\,TeV\\ }

\input{July_2008_Authors.tex}  

\affiliation{URL: http://www-cdf.fnal.gov}

\begin{abstract}
We present new limits on resonant $\tbbar$ production in $\ppbar$ collisions at $\sqrt{s}=$ 1.96\,TeV, using 1.9\,$\ifb$ of data recorded with the CDF II detector at the Fermilab Tevatron. We reconstruct a candidate mass in events with a lepton, neutrino candidate, and two or three jets, and search for anomalous $\tbbar$ production as modeled by $W' \rightarrow t\bar{b}$. We set a new limit on a right-handed $W'$ with standard model-like coupling, excluding any mass below 800\,\gevcc~at 95\%~C.L. The cross-section for any narrow, resonant $\tbbar$ production between 750 and 950\,\gevcc~is found to be less than 0.28\,pb at 95\%~C.L. We also present an exclusion of the $W'$ coupling strength versus $W'$ mass over the range 300 to 950\,\gevcc.  

\end{abstract}


\maketitle

\date{\today}

Many modifications of the standard model (SM) of particle physics include new, massive, short-lived particles with two-body decays to known fermion pairs. A classic search strategy for these states looks for resonant signals in the spectra of two-body mass distributions. Recent techniques developed to observe electroweak single-top production are well-suited to a search for unexpected $t\bar{b}$ resonances \cite{singletop}. A $t\bar{b}$ resonance (inclusion of the charge conjugate is implied throughout the text) is predicted by a wide range of models containing a massive charged vector boson, generically referred to as $W'$. The classic model is a simple extension of the SM to the left-right symmetric group $\mathrm{SU}(2)_{\mathrm{L}} \times \mathrm{SU}(2)_{\mathrm{R}} \times \mathrm{U}(1)$ \cite{lr}, which adds a right-handed charged boson $W_{\mathrm{R}}$ with universal weak coupling strength and unknown mass. The $W'$ may arise in models with other symmetry extensions: as the excitation of the $W$-boson in Kaluza-Klein extra dimensions \cite{ed}, as the techni-$\rho$ of technicolor theories \cite {tc}, or as a bosonic partner in little Higgs scenarios \cite{lh}. 

The classic limits on $W'$ are derived from searches in the $W'\rightarrow l\nu$ decay channel \cite{Wev}. For large $W'$ masses, the sensitivity in this channel is diminished by the broad Jacobian lineshapes for the lepton momentum and $W'$ transverse mass. Searches in the $t\bar{b}$ channel \cite{Wtb} avoid this difficulty and also probe models where the couplings are free parameters and the leptonic decay modes may be suppressed. Although we quantify our results using the model of a right-handed $W'$ with SM-like coupling \cite{zack}, this analysis is sensitive to any narrow state decaying to $t\bar{b}$, including {\it e.g.} a charged Higgs boson or bound states arising from new dynamics in the third generation. Searches in the $t\bar{b}$ channel complement searches for neutral states coupling to $t\bar{t}$ \cite{Z'}.

In this Letter we present a new search for an $s$-channel $W' \rightarrow \tbbar$ resonance produced in $p\bar{p}$ collisions at $\sqrt{s}=$ 1.96\,TeV at the Fermilab Tevatron. The dataset of 1.9\,$\ifb$ was recorded with the CDF II detector; a standard coordinate system \cite{coord} is used. A detailed explanation of this analysis can be found in \cite{thesis}. Our selection is based on the leptonic decay mode $t\bar{b} \rightarrow (\ell\nu b)\bar{b}$, which has been well understood in the search for electroweak single-top production \cite{singletop}. Events are expected to have a high transverse momentum ($p_{\mathrm{T}}^{~}$) electron or muon candidate, missing transverse energy (\met) from a neutrino \cite{met}, and two or three jets, at least one of which is a $b$-quark candidate. The dominant background is from $W+\mathrm{jet}$ processes and electroweak top-quark production. We reconstruct each event according to our signal hypothesis $W'\rightarrow \tbbar \rightarrow (\ell\nu b)\bar{b}$, then search the mass spectrum for a narrow resonance. If no signal is detected, we set limits on $\sigma(p\bar{p} \rightarrow W') \times \mathrm{BR}(W' \rightarrow t\bar{b})$ and on the $W'$ coupling strength $g_{W'}^{~}$.

The CDF II detector \cite{CDF} is a cylindrically-symmetric general-purpose detector. Precision charged-particle tracking is accomplished by layers of silicon microstrip detectors surrounded by a large open-cell drift chamber within a 1.4\,T solenoidal magnetic field. Outside the magnet are the electromagnetic and hadronic calorimeters, steel for hadronic shielding, and an exterior layer of muon detectors. The luminosity of the $p\bar{p}$ collisions is measured using gas Cherenkov detectors at small angles. 

We select data using online selection criteria which require a high-$p_{\mathrm{T}}^{~}$ lepton or large \met~\cite{trigger}. We identify $t\bar{b} \rightarrow \ell\nu b\bar{b}$ candidates as having an electron or muon with $p_{\mathrm{T}}^{~}\geq 20$\,\gevc. We also require $\met \geq 25$\,GeV and two or three hadronic jets with $p_{\mathrm{T}}^{~}\geq 20$\,\gevc~and $\abseta$~2.8. Jets are clustered in cones of fixed radius $\Delta R \equiv \sqrt{(\Delta \eta)^2 + (\Delta \phi)^2} \leq 0.4$, and at least one jet is required to be ``$b$-tagged'', {\it i.e.} the jet contains a secondary vertex consistent with the decay of a hadron containing a $b$-quark \cite{g4topxsec}. We reduce $Z$-decays and $t\bar{t}$ contamination by excluding events with a second charged lepton. Events consistent with cosmic ray or photon interactions are also excluded. QCD multi-jet background, which does not involve a $W$ boson, is rejected with a specific set of requirements \cite{thesis}.

The primary background process is the associated production of a $W$ boson and jets with subsequent leptonic decay of the $W$ boson ($W+\mathrm{jets}$). Approximately 70\%~of our sample are $W+\mathrm{jets}$ events containing heavy flavor ($Wb\bar{b}$, $Wc\bar{c}$, $Wcj$) or incorrectly $b$-tagged light flavor (mistags). We establish the normalization of these processes from data, and estimate the fraction of the candidate events with bottom or charm flavor using the {\tt ALPGEN} Monte Carlo event generator \cite{alpgen}. The mistagging rate for light-flavor jets is estimated from inclusive generic jet data \cite{M2}. Additional backgrounds including $t\bar{t}$ pair production, $s$-channel and $t$-channel single-top production, and diboson processes ($WW$,$WZ$,$ZZ$) are modeled using the {\tt PYTHIA} Monte Carlo event generator \cite{pythia} and are normalized to the next-to-leading-order cross-sections predicted by theory. A small multi-jet background without leptonic $W$ decay (``non-$W$'') arises when a jet is misidentified as a lepton and \met~results from jet energy mismeasurement; this background is modeled using data. The predicted SM background is detailed in Table I. The uncertainties are dominated by imprecise knowledge of the heavy flavor fraction and pertain to background rate estimates only; other systematic uncertainties are discussed later. In data we observe 1362 events with two jets and 617 events with three jets.

\begin{table}[!ht]
\begin{ruledtabular}
\caption{Predicted SM background contribution with two jets and with three jets.}
\begin{tabular}{lcc}
Background & 2 Jets & 3 Jets \\
\hline
\vspace*{-0.3cm}\\
$Wb\bar{b}$ & $409.4\pm123.4$ & $125.6\pm37.9$ \\
$Wc\bar{c} + Wcj$ & $412.4\pm127.2$ & $109.3\pm33.6$ \\
Mistags	& $276.5\pm35.0$ & $82.5\pm10.7$ \\
Non-$W$ & $53.2\pm21.3$ & $17.3\pm6.9$ \\
$\ttbar$ & $126.5\pm13.4$ & $291.8\pm36.7$ \\
Single Top ($t$-channel) & $53.3\pm7.8$ & $15.7\pm2.3$ \\
Single Top ($s$-channel) & $35.4\pm5.0$ & $11.6\pm1.6$ \\
$WW+WZ+ZZ$ & $54.4\pm4.2$ & $18.4\pm1.5$ \\
$Z$+jets & $22.6\pm3.3$ & $9.3\pm1.4$ \\ 
\hline
Total BG Prediction & $1443.8\pm254.6$ & $681.6\pm83.0$ \\
\hline
Observed & 1362	& 617 \\
\end{tabular}
\end{ruledtabular}
\label{M2Table}
\end{table}

According to the proposed $W'$ hypothesis, the $W'$ mass is given by reconstructing $M_{t\bar{b}}$ from the four-momenta of the lepton, neutrino, and two jets. The unmeasured longitudinal neutrino momentum $p^{\nu}_{\mathrm{z}}$ is quadratically constrained by assigning $M_{l\nu}=M_W=$ 80.448\,\gevcc~\cite{PDG}. We assign $p^{\nu}_{\mathrm{z}}$ to the smallest real solution or to the real part of complex solutions \cite{oldsingletop}. We assume the two highest $E_{\mathrm{T}}$ jets arise from the $b$-quarks, even for the three-jet case in which the third jet has been $b$-tagged. The reconstructed $W$ is then combined with these two leading jets, corrected to reproduce parton-level energies, to form $M_{t\bar{b}}$.

Our signal model is a $W'$ with purely right-handed decays and SM-like coupling, simulated using {\tt PYTHIA}. The model assumes a top quark mass of 175\,\gevcc. The left-handed case is not considered since the consequent $W-W'$ interference has not been observed in any precision $W$ measurements. Figure \ref{MTotal} shows the $M_{t\bar{b}}$ distribution in data superimposed with the expected signal shape for a 600\,\gevcc~$W'$ produced with a total cross-section of 9\,pb ($\sim4\times$ the prediction for a $W'$ with SM-like coupling \cite{zack}). The reconstructed width of the signal is dominated by resolution effects, particularly the jet energy resolution \cite{JES} and the incorrect assignment of jets from initial or final state radiation. Our test signal is therefore applicable for any $W'$-like object whose width is small compared to the experimental resolution. The binning is chosen so that background models have a sufficient number of entries in each bin, including the overflow bin for all values above 700\,\gevcc. 

\begin{figure}[!ht]
\begin{center}
\includegraphics[width=1.0\columnwidth]{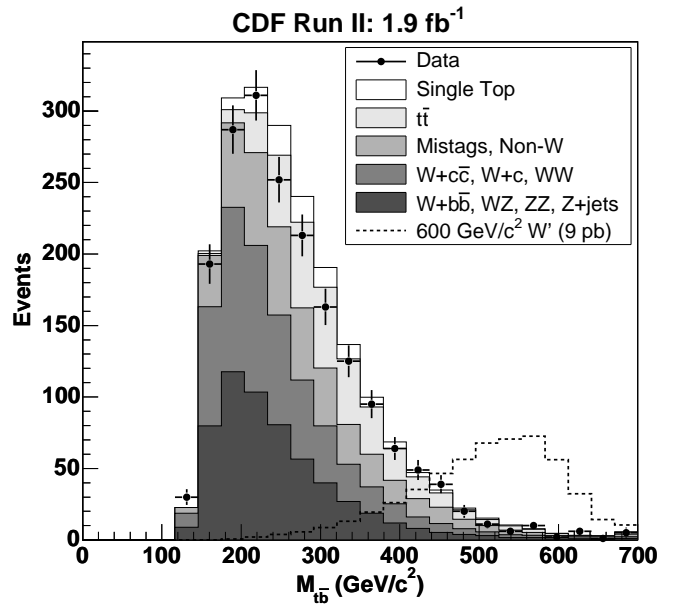}
\caption{$M_{t\bar{b}}$ for events with two jets and one $b$-tag, comparing the shapes between background and signal. Backgrounds are stacked and grouped according to similar shape. A 600\,\gevcc~$W'$ model is shown with $\sigma \times \mathrm{BR}(W' \rightarrow t\bar{b})=$ 9\,pb ($\sim4\times$ the prediction for a $W'$ with SM-like coupling).}
\label{MTotal}
\end{center}
\end{figure} 

Unlike single-top production, $W'$ production is entirely an $s$-channel process; contributions from the $t$ and $u$ channels are suppressed by the large $W'$ mass. We simulate a narrow right-handed $W'$ with SM-like coupling and a mass between 300\,\gevcc~and 950\,\gevcc~in steps of 100\,\gevcc~below 600\,\gevcc~and steps of 50\,\gevcc~above. This is the mass range to which our analysis is sensitive to changes in the signal distribution: above 950\,\gevcc~the signal events simply pile into the $M_{t\bar{b}}$ overflow bin. Since there is very little high-mass background, we are sensitive to excesses of just a few events in the tail. For $M_{W'}=$ 800\,\gevcc, our selection efficiency in the $t\bar{b}$ channel is approximately $2.8\pm1.0$\%. An excess of ten events, for example, would correspond to a Tevatron cross-section of 0.18\,pb.

The branching ratios of a right-handed $W'$ depend on whether decay to $\nu_{\mathrm{R}}^{~}$ is allowed; we consider both possibilities. If leptonic decay is forbidden, as for a leptophobic $W'$ or when $M_{W'}<M_{\nu_{\mathrm{R}}^{~}}$, the $M_{t\bar{b}}$ prediction simply has a slightly larger normalization. For example, if $M_{W'}=$ 800\,\gevcc, $\sigma \times \mathrm{BR}(W'\rightarrow t\bar{b})$ is predicted to be 0.337\,pb if leptonic decays are forbidden and 0.262\,pb if they are allowed.

We set frequentist limits on $W'\rightarrow t\bar{b}$ using the measure $CL_s$ from \cite{CLs}, which is defined as the probability of background plus a specified signal fraction matching the data ($P_{\mathrm{S+B}}$) divided by the probability of a background-only model matching the data ($P_{\mathrm{B}}$). Sources of uncertainty are treated using a large series of trials ($\sim$50k) for both cases. Each trial is produced by randomly varying all uncertain parameters in the model prediction within a Gaussian constraint about their nominal values. $P_{\mathrm{S+B}}$ is determined from the fraction of the S+B trials with a minimized $\Delta \chi^2 = \chi^2(\mathrm{Data}|\mathrm{S+B})-\chi^2(\mathrm{Data}|\mathrm{B})$ larger than in data; $P_{\mathrm{B}}$ is analogous. The 95\% C.L. limit is set by adjusting the signal fraction assumed in the S+B model until $CL_{\mathrm{s}}=0.05$. 

Our event selection introduces various sources of systematic uncertainty. These are manifest as errors in both the rates and shapes the mass distributions for our signal and background models. They include: jet-energy scale (JES), $b$-tagging efficiencies, lepton identification and trigger efficiencies, recorded luminosity, quantity of initial and final state radiation, parton distribution functions, factorization and renormalization scale, and MC modeling. Our limit procedure evaluates their impact by making reasonable variations in the model parameters and re-simulating the analysis \cite{thesis}. 

The systematic uncertainties are dominated by JES and the $b$-tagging rate uncertainties for the signal. JES uncertainty is modeled by calculating 1$\sigma$ shifts in each jet-energy correction and adding the results in quadrature. The uncertainty in $b$-tagging efficiency is determined by binning the $b$-tagging rate as a function of energy for multi-jet data. The uncertainty is found to be proportional to the jet energy, allowing extrapolation to the higher energies common for our $W'$ signal. This jet-energy weighted uncertainty on the $b$-tagging rate leads to acceptance errors as large as 40\%~for a 950\,\gevcc~$W'$. Including all such sources of uncertainty in our model results in the expected upper limit on the cross-section increasing by 30-40\%.

Applying the full limit procedure, we set 95\%~C.L. upper limits on $\sigma \times \mathrm{BR}(W' \rightarrow t\bar{b})$ as listed in Table II for a right-handed $W'$ with SM-like coupling. Predicted cross-sections for such a $W'$ \cite{zack} are shown in Figure \ref{Limit_obs}: we set new 95\%~C.L. limits of $M_{W'}>$ 800\,\gevcc~including leptonic decays, and $M_{W'}>$ 825\,\gevcc~if leptonic decays are forbidden. The best prior result used 0.9\,$\ifb$ and found $M_{W'} \geq$ 768\,\gevcc~if leptonic decays are forbidden \cite{Wtb}. These results are quoted for a top quark mass of 175\,\gevcc~and thus are slightly conservative: using the smaller world-average would increase the $t\bar{b}$ branching fraction.

\begin{table}[!ht]
\begin{ruledtabular}
\caption{95\% C.L. limits on $\sigma \times \mathrm{BR}(W' \rightarrow t\bar{b})$ as function of $M_{W'}$ for a right-handed $W'$ with SM-like coupling. The expected limit is quoted with the range of values into which our observation should fall 68\% of the time assuming no signal is present.}
\begin{tabular}{ccc}
${M_{W'}~(\gevcc)}$ & Expected Limit (pb) & Observed Limit (pb)\\
\hline
\vspace*{-0.3cm}\\
300 & $1.56_{-0.45}^{+0.62}$ & 1.59 \\
400 & $1.04_{-0.30}^{+0.44} $ & 1.17\\
500 & $0.74_{-0.22}^{+0.35}$ & 0.84\\
600 & $0.54_{-0.17}^{+0.24}$ & 0.44\\
650 & $0.46_{-0.13}^{+0.21}$ & 0.39\\
700 & $0.40_{-0.12}^{+0.17}$ & 0.32\\
750 & $0.33_{-0.09}^{+0.15}$ & 0.28\\
800 & $0.30_{-0.09}^{+0.13}$ & 0.26\\
850 & $0.28_{-0.08}^{+0.13}$ & 0.25\\
900 & $0.28_{-0.08}^{+0.13} $ & 0.26\\
950 & $0.30_{-0.09}^{+0.13} $ & 0.28\\
\end{tabular}
\end{ruledtabular}
\label{LimitTable}
\end{table}

\begin{figure}[!ht]
\begin{center}
\includegraphics[width=1.0\columnwidth]{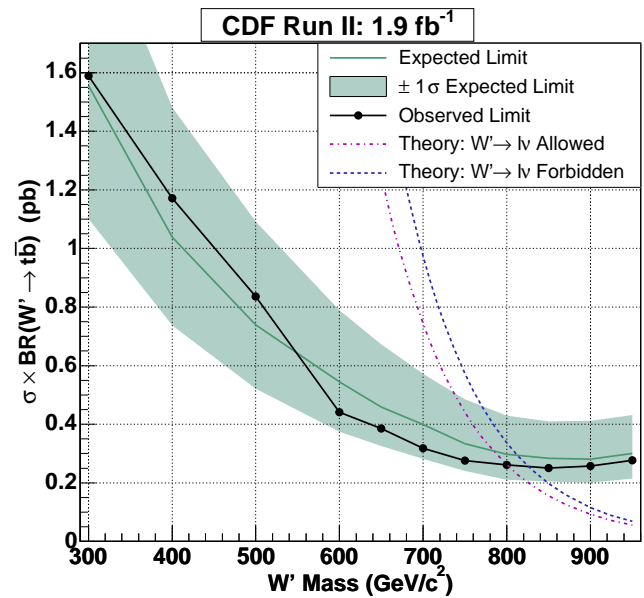}
\caption{Expected and observed 95\%~C.L. limits on $\sigma \times \mathrm{BR}(W' \rightarrow t\bar{b})$ as function of $M_{W'}$ for 1.9\,$\ifb$, along with theoretical predictions. A right-handed $W'$ with SM-like couplings is excluded for $W'$ masses below 800\,\gevcc.}
\label{Limit_obs}
\end{center} 
\end{figure}  

For a simple $s$-channel model with effective coupling $g_{W'}^{~}$, the cross-section is proportional to $g_{W'}^4$. Relaxing the assumption of the universal weak coupling, our cross-section limits can be rewritten as upper limits on $g_{W'}^{~}$ as a function of $M_{W'}$. The excluded region of the $g_{W'}^{~}-M_{W'}$ plane is shown in Figure \ref{Coupling_obs}, with $g_{W'}^{~}$ in units of $g_{W}^{~}$. At $M_{W'} = 300~\gevcc$, we limit (95\% C.L.) the effective coupling to be less than 0.40 of the $W$ boson coupling. In this more general case, the effective cross-section for any narrow, resonant $\tbbar$ production between 750 and 950\,\gevcc~is found to be less than 0.28\,pb at 95\%~C.L. 

\begin{figure}[!ht]
\begin{center}
\includegraphics[width=1.0\columnwidth]{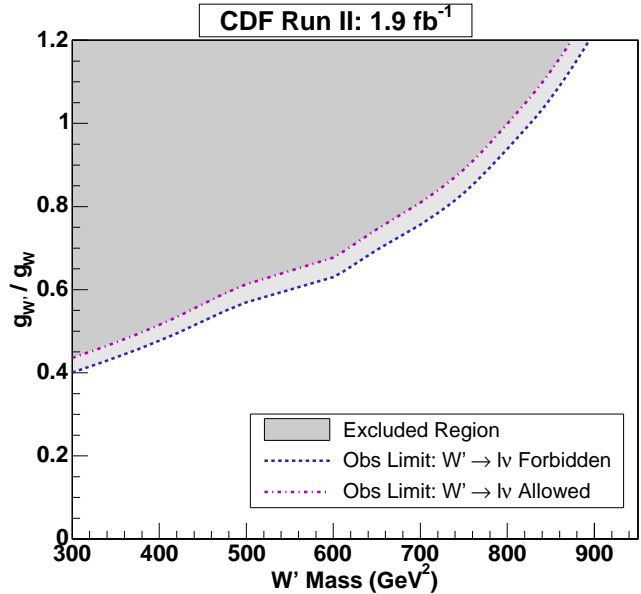}
\caption{Observed 95\% C.L. limits on the coupling strength of a right-handed $W'$ compared to the SM $W$ boson coupling, $g_{W'}^{~}$/$g_{W}^{~}$, as function of $M_{W'}$ for 1.9\,$\ifb$. The shaded region above the dashed lines are excluded.}
\label{Coupling_obs}
\end{center} 
\end{figure}

\begin{acknowledgments}
We thank the Fermilab staff and the technical staffs of the participating institutions for their vital contributions. This work was supported by the U.S. Department of Energy and National Science Foundation; the Italian Istituto Nazionale di Fisica Nucleare; the Ministry of Education, Culture, Sports, Science and Technology of Japan; the Natural Sciences and Engineering Research Council of Canada; the National Science Council of the Republic of China; the Swiss National Science Foundation; the A.P. Sloan Foundation; the Bundesministerium f\"ur Bildung und Forschung, Germany; the Korean Science and Engineering Foundation and the Korean Research Foundation; the Science and Technology Facilities Council and the Royal Society, UK; the Institut National de Physique Nucleaire et Physique des Particules/CNRS; the Russian Foundation for Basic Research; the Ministerio de Ciencia e Innovaci\'{o}n, and Programa Consolider-Ingenio 2010, Spain; the Slovak R\&D Agency; and the Academy of Finland.
\end{acknowledgments}

\newpage




\end{document}

%% file: July_2008_Authors.tex
\affiliation{Institute of Physics, Academia Sinica, Taipei, Taiwan 11529, Republic of China} 
\affiliation{Argonne National Laboratory, Argonne, Illinois 60439} 
\affiliation{University of Athens, 157 71 Athens, Greece} 
\affiliation{Institut de Fisica d'Altes Energies, Universitat Autonoma de Barcelona, E-08193, Bellaterra (Barcelona), Spain} 
\affiliation{Baylor University, Waco, Texas  76798} 
\affiliation{Istituto Nazionale di Fisica Nucleare Bologna, $^y$University of Bologna, I-40127 Bologna, Italy} 
\affiliation{Brandeis University, Waltham, Massachusetts 02254} 
\affiliation{University of California, Davis, Davis, California  95616} 
\affiliation{University of California, Los Angeles, Los Angeles, California  90024} 
\affiliation{University of California, San Diego, La Jolla, California  92093} 
\affiliation{University of California, Santa Barbara, Santa Barbara, California 93106} 
\affiliation{Instituto de Fisica de Cantabria, CSIC-University of Cantabria, 39005 Santander, Spain} 
\affiliation{Carnegie Mellon University, Pittsburgh, PA  15213} 
\affiliation{Enrico Fermi Institute, University of Chicago, Chicago, Illinois 60637}
\affiliation{Comenius University, 842 48 Bratislava, Slovakia; Institute of Experimental Physics, 040 01 Kosice, Slovakia} 
\affiliation{Joint Institute for Nuclear Research, RU-141980 Dubna, Russia} 
\affiliation{Duke University, Durham, North Carolina  27708} 
\affiliation{Fermi National Accelerator Laboratory, Batavia, Illinois 60510} 
\affiliation{University of Florida, Gainesville, Florida  32611} 
\affiliation{Laboratori Nazionali di Frascati, Istituto Nazionale di Fisica Nucleare, I-00044 Frascati, Italy} 
\affiliation{University of Geneva, CH-1211 Geneva 4, Switzerland} 
\affiliation{Glasgow University, Glasgow G12 8QQ, United Kingdom} 
\affiliation{Harvard University, Cambridge, Massachusetts 02138} 
\affiliation{Division of High Energy Physics, Department of Physics, University of Helsinki and Helsinki Institute of Physics, FIN-00014, Helsinki, Finland} 
\affiliation{University of Illinois, Urbana, Illinois 61801} 
\affiliation{The Johns Hopkins University, Baltimore, Maryland 21218} 
\affiliation{Institut f\"{u}r Experimentelle Kernphysik, Universit\"{a}t Karlsruhe, 76128 Karlsruhe, Germany} 
\affiliation{Center for High Energy Physics: Kyungpook National University, Daegu 702-701, Korea; Seoul National University, Seoul 151-742, Korea; Sungkyunkwan University, Suwon 440-746, Korea; Korea Institute of Science and Technology Information, Daejeon, 305-806, Korea; Chonnam National University, Gwangju,500-757, Korea; Chonbuk National University, Jeonju 561-756, Korea} 
\affiliation{Ernest Orlando Lawrence Berkeley National Laboratory, Berkeley, California 94720} 
\affiliation{University of Liverpool, Liverpool L69 7ZE, United Kingdom} 
\affiliation{University College London, London WC1E 6BT, United Kingdom} 
\affiliation{Centro de Investigaciones Energeticas Medioambientales y Tecnologicas, E-28040 Madrid, Spain} 
\affiliation{Massachusetts Institute of Technology, Cambridge, Massachusetts  02139} 
\affiliation{Institute of Particle Physics: McGill University, Montr\'{e}al, Qu\'{e}bec, Canada H3A~2T8; Simon Fraser University, Burnaby, British Columbia, Canada V5A~1S6; University of Toronto, Toronto, Ontario, Canada M5S~1A7; and TRIUMF, Vancouver, British Columbia, Canada V6T~2A3} 
\affiliation{University of Michigan, Ann Arbor, Michigan 48109} 
\affiliation{Michigan State University, East Lansing, Michigan  48824}
\affiliation{Institution for Theoretical and Experimental Physics, ITEP, Moscow 117259, Russia} 
\affiliation{University of New Mexico, Albuquerque, New Mexico 87131} 
\affiliation{Northwestern University, Evanston, Illinois  60208} 
\affiliation{The Ohio State University, Columbus, Ohio  43210} 
\affiliation{Okayama University, Okayama 700-8530, Japan} 
\affiliation{Osaka City University, Osaka 588, Japan} 
\affiliation{University of Oxford, Oxford OX1 3RH, United Kingdom} 
\affiliation{Istituto Nazionale di Fisica Nucleare, Sezione di Padova-Trento, $^z$University of Padova, I-35131 Padova, Italy} 
\affiliation{LPNHE, Universite Pierre et Marie Curie/IN2P3-CNRS, UMR7585, Paris, F-75252 France} 
\affiliation{University of Pennsylvania, Philadelphia, Pennsylvania 19104}
\affiliation{Istituto Nazionale di Fisica Nucleare Pisa, $^{aa}$University of Pisa, $^{bb}$University of Siena and $^{cc}$Scuola Normale Superiore, I-56127 Pisa, Italy} 
\affiliation{University of Pittsburgh, Pittsburgh, Pennsylvania 15260} 
\affiliation{Purdue University, West Lafayette, Indiana 47907} 
\affiliation{University of Rochester, Rochester, New York 14627} 
\affiliation{The Rockefeller University, New York, New York 10021} 
\affiliation{Istituto Nazionale di Fisica Nucleare, Sezione di Roma 1, $^{dd}$Sapienza Universit\`{a} di Roma, I-00185 Roma, Italy} 

\affiliation{Rutgers University, Piscataway, New Jersey 08855} 
\affiliation{Texas A\&M University, College Station, Texas 77843} 
\affiliation{Istituto Nazionale di Fisica Nucleare Trieste/Udine, I-34100 Trieste, $^{ee}$University of Trieste/Udine, I-33100 Udine, Italy} 
\affiliation{University of Tsukuba, Tsukuba, Ibaraki 305, Japan} 
\affiliation{Tufts University, Medford, Massachusetts 02155} 
\affiliation{Waseda University, Tokyo 169, Japan} 
\affiliation{Wayne State University, Detroit, Michigan  48201} 
\affiliation{University of Wisconsin, Madison, Wisconsin 53706} 
\affiliation{Yale University, New Haven, Connecticut 06520} 
\author{T.~Aaltonen}
\affiliation{Division of High Energy Physics, Department of Physics, University of Helsinki and Helsinki Institute of Physics, FIN-00014, Helsinki, Finland}
\author{J.~Adelman}
\affiliation{Enrico Fermi Institute, University of Chicago, Chicago, Illinois 60637}
\author{T.~Akimoto}
\affiliation{University of Tsukuba, Tsukuba, Ibaraki 305, Japan}
\author{B.~\'{A}lvarez~Gonz\'{a}lez$^t$}
\affiliation{Instituto de Fisica de Cantabria, CSIC-University of Cantabria, 39005 Santander, Spain}
\author{S.~Amerio$^z$}
\affiliation{Istituto Nazionale di Fisica Nucleare, Sezione di Padova-Trento, $^z$University of Padova, I-35131 Padova, Italy} 

\author{D.~Amidei}
\affiliation{University of Michigan, Ann Arbor, Michigan 48109}
\author{A.~Anastassov}
\affiliation{Northwestern University, Evanston, Illinois  60208}
\author{A.~Annovi}
\affiliation{Laboratori Nazionali di Frascati, Istituto Nazionale di Fisica Nucleare, I-00044 Frascati, Italy}
\author{J.~Antos}
\affiliation{Comenius University, 842 48 Bratislava, Slovakia; Institute of Experimental Physics, 040 01 Kosice, Slovakia}
\author{G.~Apollinari}
\affiliation{Fermi National Accelerator Laboratory, Batavia, Illinois 60510}
% ...

\collaboration{CDF Collaboration\footnote{With visitors from $^a$University of Massachusetts Amherst, Amherst, Massachusetts 01003,
$^b$Universiteit Antwerpen, B-2610 Antwerp, Belgium, 
$^c$University of Bristol, Bristol BS8 1TL, United Kingdom,
$^d$Chinese Academy of Sciences, Beijing 100864, China, 
$^e$Istituto Nazionale di Fisica Nucleare, Sezione di Cagliari, 09042 Monserrato (Cagliari), Italy,
$^f$University of California Irvine, Irvine, CA  92697, 
$^g$University of California Santa Cruz, Santa Cruz, CA  95064, 
$^h$Cornell University, Ithaca, NY  14853, 
$^i$University of Cyprus, Nicosia CY-1678, Cyprus, 
$^j$University College Dublin, Dublin 4, Ireland,
$^k$University of Edinburgh, Edinburgh EH9 3JZ, United Kingdom, 
$^l$University of Fukui, Fukui City, Fukui Prefecture, Japan 910-0017
$^m$Kinki University, Higashi-Osaka City, Japan 577-8502
$^n$Universidad Iberoamericana, Mexico D.F., Mexico,
$^o$University of Iowa, Iowa City, IA  52242,
$^p$Queen Mary, University of London, London, E1 4NS, England,
$^q$University of Manchester, Manchester M13 9PL, England, 
$^r$Nagasaki Institute of Applied Science, Nagasaki, Japan, 
$^s$University of Notre Dame, Notre Dame, IN 46556,
$^t$University de Oviedo, E-33007 Oviedo, Spain, 
$^u$Texas Tech University, Lubbock, TX  79609, 
$^v$IFIC(CSIC-Universitat de Valencia), 46071 Valencia, Spain,
$^w$University of Virginia, Charlottesville, VA  22904,
$^x$Bergische Universit\"at Wuppertal, 42097 Wuppertal, Germany,
$^{ff}$On leave from J.~Stefan Institute, Ljubljana, Slovenia, 
}}
\noaffiliation

%% file: prl5.bbl
\begin{thebibliography}{99}
\bibitem{coord} We use coordinates where $\phi$ is the azimuthal angle, $\theta$ is the polar angle with respect to the proton beam axis, transverse energy is $E_{\mathrm{T}} = E\sin(\theta)$, and the pseudorapidity is $\rm{\eta = -ln[tan(\theta/2)]}$.
\bibitem{met} Missing transverse energy, $\met$, is defined as the magnitude of the vector $ -\sum_i E_{\mathrm{T}}^i \vec n_i$ where $E_{\mathrm{T}}^i$ are the magnitudes of transverse energy contained in each calorimeter tower $i$, and $\vec n_i$ is the unit vector from the interaction vertex to the tower in the transverse ($x,y$) plane.
\end{thebibliography}
